\documentclass[11pt,a4paper]{article}
\usepackage[english]{babel}
\usepackage{amsmath,amsthm,amssymb,epsfig,latexsym}
\usepackage{color}
\usepackage{ulem}
\usepackage[numbers,square,sort&compress]{natbib}

\newcommand{\so}{\scriptscriptstyle \rm I}
\newcommand{\st}{\scriptscriptstyle \rm I\hspace{-1pt}I}

\newcommand{\la}{u}

\newcommand{\muc}{v^{\scriptscriptstyle C}}
\newcommand{\mub}{v^{\scriptscriptstyle B}}

\newcommand{\bla}{\bar u}
\newcommand{\bmu}{\bar v}
\newcommand{\blac}{\bar{u}^{\scriptscriptstyle C}}
\newcommand{\blab}{\bar{u}^{\scriptscriptstyle B}}
\newcommand{\bmuc}{\bar{v}^{\scriptscriptstyle C}}
\newcommand{\bmub}{\bar{v}^{\scriptscriptstyle B}}
\newcommand{\blacb}{\bar{u}^{\scriptscriptstyle C,B}}
\newcommand{\bmucb}{\bar{v}^{\scriptscriptstyle C,B}}
\newcommand{\blacp}{\bar{u'}^{\scriptscriptstyle C}}
\newcommand{\blabp}{\bar{u'}^{\scriptscriptstyle B}}
\newcommand{\bmucp}{\bar{v'}^{\scriptscriptstyle C}}
\newcommand{\bmubp}{\bar{v'}^{\scriptscriptstyle B}}

\newcommand{\bxi}{\bar\xi}

\newcommand{\cc}{\varkappa}

\newcommand{\ZMRp}[1]{T^{(1;\text{\tiny R})}_{#1}[0]}
\newcommand{\ZMLp}[1]{T^{(1;\text{\tiny L})}_{#1}[0]}
\newcommand{\be}[1]{\begin{equation}\label{#1}}
\newcommand{\ba}[1]{\begin{multline}\label{#1}}
\newcommand{\ee}{\end{equation}}
\newcommand{\ea}{\end{eqnarray}}

\newcommand{\num}{\\\rule{0pt}{20pt}}

\newcommand{\tr}{\mathop{\rm tr}}

\newtheorem{thm}{Theorem}[section]
\newtheorem{prop}{Proposition}[section]
\newtheorem{lemma}{Lemma}[section]

 \makeatletter
 \@addtoreset{equation}{section}
 \makeatother
 
\newcommand{\bea}{\begin{eqnarray}}
\newcommand{\eea}{\end{eqnarray}}

\begin{document}

\begin{flushright}
LAPTH-010/15
\end{flushright}

\vspace{22pt}

\begin{center}
\begin{LARGE}
{\bf Form factors of local operators\\ in a one-dimensional two-component Bose gas}
\end{LARGE}

\vspace{40pt}

\begin{large}
{S.~Pakuliak${}^a$, E.~Ragoucy${}^b$, N.~A.~Slavnov${}^c$\footnote{pakuliak@jinr.ru, eric.ragoucy@lapth.cnrs.fr, nslavnov@mi.ras.ru}}
\end{large}

 \vspace{12mm}

\vspace{4mm}

${}^a$ {\it Laboratory of Theoretical Physics, JINR, 141980 Dubna, Moscow reg., Russia,\\
Moscow Institute of Physics and Technology, 141700, Dolgoprudny, Moscow reg., Russia,\\
Institute of Theoretical and Experimental Physics, 117259 Moscow, Russia}

\vspace{4mm}

${}^b$ {\it Laboratoire de Physique Th\'eorique LAPTH, CNRS and Universit\'e de Savoie,\\
BP 110, 74941 Annecy-le-Vieux Cedex, France}

\vspace{4mm}

${}^c$ {\it Steklov Mathematical Institute,
Moscow, Russia}

\end{center}


\vspace{4mm}


\begin{abstract}
We consider a one-dimensional model of a two-component Bose gas and study form factors of local operators
in this model. For this aim we use an approach based on the algebraic Bethe ansatz.
We show that the form factors under consideration can be reduced to those of the monodromy matrix entries
in a generalized $GL(3)$-invariant model. In this way we derive determinant representations for the form factors
of local operators.
\end{abstract}

\vspace{1cm}

\vspace{2mm}

\section{Introduction}

In this paper we  consider a one-dimensional model of  Two-Component Bose Gas with $\delta$-function repulsive interaction (TCBG model).
This model is a generalization of the Lieb--Liniger model \cite{LiebL63,Lieb63}, in which Bose fields
have two internal degrees of freedom.  The  version with one internal degree of freedom
is also known under different denominations:  the Quantum nonlinear Schr\"odinger equation, Tonks--Girardeau gas \cite{Tonks36,Gir60} or Gross--Pitaevskii model \cite{Gross61,Pit61}, so that one can say that we are studying a two-component version of these models.
The TCBG model was solved by C.~N.~Yang \cite{Yang67} who have found
the eigenvectors and the spectrum of the Hamiltonian. The general approach to the solution of the
model with $n$ internal degrees of freedom (multi-component Bose gas) was given in \cite{Sath68} (see also
\cite{Sath75,Gaud83}). The nested algebraic Bethe ansatz was applied to this model in \cite{Kul81,KulRes82}.

We consider the TCBG model on a finite interval $[0,L]$ with periodic boundary conditions. The Hamiltonian of the model has the form
 \be{HamQ}
 H=\int_0^L\left(\partial_x\Psi^\dagger_\alpha\partial_x\Psi_\alpha+\cc\Psi^\dagger_\alpha\Psi^\dagger_\beta\Psi_\beta\Psi_\alpha\right)\,dx,
 \ee
where $\cc>0$ is a coupling constant, $\alpha,\beta=1,2$ and the summation over repeated subscripts is assumed.
The Bose fields $\Psi_\alpha(x)$ and $\Psi_\alpha^\dagger(x)$ satisfy canonical commutation relations
\be{com-con}
[\Psi_\alpha(x),\Psi_\beta^\dagger(y)]=\delta_{\alpha\beta}\delta(x-y).
\ee

The basis in the Fock space of the model is constructed by acting with operators
$\Psi_\alpha^\dagger(x)$ onto the Fock vacuum $|0\rangle$ defined as
\be{vac-def}
\Psi_\alpha(x)|0\rangle=0,\qquad \langle 0|\Psi_\alpha^\dagger(x)=0, \qquad \langle 0|0\rangle=1.
\ee

Form factors of local operators of the TCBG model were studied in \cite{PozOK12}. There, determinant representations for the form factors
were obtained in some particular cases of the Hamiltonian eigenstates. In the present paper we consider a general case of form factors and
obtain determinant representations for them. Our approach is based on recent results obtained in \cite{PakRS15c}. There we developed a method of calculating form factors of local operators in a wide class of $GL(3)$-invariant integrable models solvable by the nested algebraic Bethe ansatz. However the method developed
in \cite{PakRS15c} can be applied to the TCBG model only partly. Calculation of some form factors needs certain modifications. We consider these questions in
the present paper.

In the models considered  in \cite{PakRS15c} form factors of all monodromy matrix elements can be obtained by
the zero modes method \cite{PakRS15a} from an initial form factor corresponding to
one of the monodromy matrix entries.
In its turn, the initial form factor should be calculated straightforwardly. The calculation is based on a sum formula for the scalar product of Bethe vectors \cite{Res86} and a summation identity
(see \eqref{Wpart}). However the choice of the initial form factor is not fixed  and we can use this freedom to deduce new summation identities. In \cite{PakRS15c} we have chosen as the initial form factor the
one of a diagonal element of the monodromy matrix. Making another choice for this initial form factor, one can redo the full process through minor modifications of our calculations. Then, comparing the two final results, we obtain new summation identities. The  latter can be directly used for the calculating form factors in the TCBG model.

The paper is organized as follows. In section~\ref{S-ABATCBG} we describe  the general settings of the algebraic Bethe ansatz. In particular,
 we introduce a composite model and give definitions of the zero modes in  the TCBG model.
In section~\ref{S-MR} we formulate the main results of the paper and partly prove them. In particular we show that the determinant representations for the operators
$\Psi^\dagger_i(x)\Psi_j(x)$ directly follow from the results of \cite{PakRS15c}. We also show that  the form factors of Bose fields
are related one to each other by simple transformations.
In   section~\ref{S-SI} we derive a determinant representation for the form factor of the field $\Psi_2(x)$.
In appendix we prove a summation identity that we use in section~\ref{S-SI}.

\section{General scheme of the algebraic Bethe ansatz\label{S-ABATCBG}}

In this section we describe a general scheme of the nested algebraic Bethe ansatz for $GL(3)$-invariant models.
We also point out a specification of some parameters in the case of the TCBG model.

The $GL(3)$-invariant models are described by the following $R$-matrix
acting on a space $V_1\otimes V_2$, with $V_j=\mathbb{C}^3$
 \be{R-mat}
 R(u,v)=\mathbf{I}+g(u,v)\mathbf{P},\qquad g(u,v)=\frac{c}{u-v}.
 \ee
Here $\mathbf{I}$ is the identity matrix in $V_1\otimes V_2$, $\mathbf{P}$ is the permutation matrix
that exchanges $V_1$ and $V_2$. The parameter $c$ is related to the coupling constant of the TCBG model by $c=-i\cc$.

The monodromy matrix $T(u)$ satisfies a standard $RTT$-relation
\be{RTT}
R_{12}(u,v)T_1(u)T_2(v)=T_2(v)T_1(u)R_{12}(u,v).
\ee
The monodromy matrix $T(u)$ acts in $\mathbb{C}^3\otimes\mathcal{H}$. In the TCBG model
$\mathcal{H}$ is the Fock space of the Hamiltonian \eqref{HamQ}. Equation \eqref{RTT} holds in the tensor product $V_1\otimes V_2\otimes\mathcal{H}$, and the  matrices $T_k(w)$ act non-trivially in
$V_k\otimes \mathcal{H}$.
The Fock vacuum vector $|0\rangle$ is annihilated by the operators $T_{ij}(w)$ with $i>j$. The dual vector $\langle0|$ is annihilated by the operators $T_{ij}(w)$ with $i<j$. Both vectors are eigenvectors of the diagonal entries of the monodromy matrix
 \be{ratios}
 T_{kk}(w)|0\rangle=r_k(w)|0\rangle, \qquad   \langle0|T_{kk}(w)=r_k(w)\langle0|,\quad k=1,2,3.
 \ee
Without loss of generality we assume that $r_2(w)=1$. In the TCBG model we also have $r_1(w)=1$ and $r_3(w)=e^{iwL}$.
However, up to a certain point it is convenient not to use explicit expressions for the functions $r_k(w)$. Therefore we shall
continue to use the notation $r_k(w)$, making its specification if necessary.

Bethe vectors are certain polynomials in operators $T_{ij}(u)$ with $i<j$ acting on the  vector
$|0\rangle$ \cite{KulR83,TarVar93,KhoPakT07,KhoPak08,BelPRS12c}. In the $GL(3)$-invariant models they depend on two sets of variables called Bethe parameters.
We denote the Bethe vectors $\mathbb{B}_{a,b}(\bla;\bmu)$. Here the Bethe parameters are $\bla=\{u_1,\dots,u_a\}$ and
$\bmu=\{v_1,\dots,v_b\}$. We call them $u$-type variables and $v$-type variables. The subscripts $a$ and $b$ ($a,b=0,1,\dots$) respectively denote the cardinalities of the sets $\bla$ and $\bmu$.
The peculiarity  of the TCBG model is that the Bethe vectors do not depend on the operators $T_{12}(u)$: they are polynomials
in $T_{13}(v)$ and $T_{23}(v)$ only. One more restriction for the Bethe vectors of the TCBG model is that $a\le b$. However these
conditions do not play an essential role in our considerations.

Similarly we can construct dual Bethe vectors in the dual space as polynomials in operators $T_{ij}(u)$ with $i>j$ acting on
$\langle0|$. We denote them $\mathbb{C}_{a,b}(\bla;\bmu)$ with the same meaning for the arguments and
subscripts. Dual Bethe vectors  of the TCBG model exist for $a\le b$ and they do not depend on $T_{21}(u)$.

\subsection{Notation\label{SS-Not}}

Besides the function $g(u,v)$ we also introduce a function $f(u,v)$
\be{univ-not}
 f(u,v)=\frac{u-v+c}{u-v}.
\ee

We denote sets of variables by bar:  $\bla$, $\bmu$ etc. If necessary, the cardinalities of the sets are
given in special comments.
Individual elements of the sets are denoted by subscripts: $w_j$, $\la_k$ etc.
We say that $\bar x=\bar x'$,
if $\#\bar x=\#\bar x'$ and $x_i=x'_i$ (up to a permutation) for $i=1,\dots,\#\bar x$. We say that $\bar x\ne \bar x'$ otherwise.

Below we consider partitions of sets into subsets. The notation $\bar u\Rightarrow\{\bla_{\so},\bla_{\st}\}$ means that the set $\bla$
is divided into two disjoint subsets $\bla_{\so}$ and $\bla_{\st}$. As a rule, we use roman numbers for subscripts of subsets: $\bla_{\so}$, $\bmu_{\rm ii}$ etc.
However, if we deal with a big quantity of subsets, then we use standard arabic numbers for their notation. In such cases we explicitly indicate it.

Similarly to the paper \cite{PakRS15b} we use a shorthand notation for products of  some functions.
Namely, if the functions $r_i$ \eqref{ratios} or the function $f$ \eqref{univ-not} depend
on sets of variables, this means that one should take the product over the corresponding set.
For example,
 \be{SH-prod}
 r_1(\bla)=\prod_{\la_k\in\bla} r_1(\la_k);\quad
 f(z, \bar w)= \prod_{w_j\in\bar w} f(z, w_j);\quad
 f(\bmu_{\st},\bla_{\so})=\prod_{u_j\in\bla_{\so}}\prod_{v_k\in\bmu_{\st}} f(v_k,u_j).
 \ee
In the last equation it is assumed that the sets $\bla$ and $\bmu$ are divided into several subsets and the product is taken over the
subsets $\bmu_{\st}$ and $\bla_{\so}$. By definition any product with respect to the empty set is equal to $1$. If we have a double product, then it is also equal to $1$ if at least
one of sets is empty.

In section~\ref{SS-CM} we shall introduce several new scalar functions and will extend
the convention \eqref{SH-prod} to their products.

\subsection{On-shell Bethe vectors\label{SS-OBV}}

In the algebraic Bethe
ansatz the role of a quantum Hamiltonian is played by  the transfer matrix. It is the trace in the auxiliary space of the monodromy matrix: $\tr T(u)$.
The eigenstates of the transfer matrix are called {\it on-shell} Bethe vectors. The eigenstates of the transfer matrix in the dual space are called
dual on-shell Bethe vectors\footnote{%
For simplicity here and below we do not distinguish between vectors and dual vectors, because their properties are completely analogous to each other.}.

A (dual) Bethe vector becomes on-shell, if the Bethe parameters satisfy the system of Bethe equations.
We give this system in a slightly unusual form
\be{AEigenS-1}
r_1(\bla_{\so})=\frac{f(\bla_{\so},\bla_{\st})}{f(\bla_{\st},\bla_{\so})}f(\bmu,\bla_{\so}),\qquad
r_3(\bmu_{\so})=\frac{f(\bmu_{\st},\bmu_{\so})}{f(\bmu_{\so},\bmu_{\st})}f(\bmu_{\so},\bla).
\ee
These equations should hold for arbitrary partitions of the sets $\bla$ and $\bmu$ into subsets
$\{\bla_{\so},\;\bla_{\st}\}$ and $\{\bmu_{\so},\;\bmu_{\st}\}$ with $\#\bla_{\so}=\#\bmu_{\so}=1$. It is easy to see that
if the sets $\bla$ and $\bmu$ satisfy the system \eqref{AEigenS-1}, then they satisfy the same system without the
restriction $\#\bla_{\so}=\#\bmu_{\so}=1$.  In particular, if $\bla_{\st}=\bmu_{\st}=\emptyset$, we obtain
\be{prod-r}
r_1(\bla)=r_3(\bmu)=f(\bmu,\bla)=1,
\ee
because $r_1(u)=1$ in the TCBG model.

If the sets $\bla$ and $\bmu$ satisfy \eqref{AEigenS-1}, then
\be{T-Action}
\tr T(w) \mathbb{B}_{a,b}(\bla,\bmu)= \tau(w|\bla,\bmu)\mathbb{B}_{a,b}(\bla,\bmu),\qquad
 \mathbb{C}_{a,b}(\bla,\bmu)\tr T(w)= \tau(w|\bla,\bmu)\mathbb{C}_{a,b}(\bla,\bmu),
 \ee
with
\be{tau-def}
\tau(w|\bla,\bmu)=r_1(w)f(\bla,w)+f(w,\bla)f(\bmu,w)+r_3(w)f(w,\bmu).
\ee

\subsection{Scalar products and form factors}

Scalar product of generic Bethe vectors is defined as follows:
\be{SP-def}
 \mathcal{S}_{a,b}\equiv \mathcal{S}_{a,b}(\blac,\bmuc;\blab,\bmub)= \mathbb{C}_{a,b}(\blac;\bmuc)\mathbb{B}_{a,b}(\blab;\bmub).
 \ee
An expression for the scalar products in terms of a sum over partitions of Bethe parameters (sum formula) was found in \cite{Res86}
 \begin{multline}\label{Resh-SP}
 \mathcal{S}_{a,b}=\sum r_1(\blab_{\so})r_1(\blac_{\st})
 r_3(\bmuc_{\st})r_3(\bmub_{\so})
  f(\blac_{\so},\blac_{\st})  f(\blab_{\st},\blab_{\so})     \num
 \times f(\bmuc_{\st},\bmuc_{\so})   f(\bmub_{\so},\bmub_{\st})\frac{f(\bmuc_{\so},\blac_{\so})f(\bmub_{\st},\blab_{\st})}{f(\bmuc,\blac)f(\bmub,\blab)}
  \;Z_{a_{\st},b_{\so}}(\blac_{\st};\blab_{\st}|\bmuc_{\so};\bmub_{\so})
 Z_{a_{\so},b_{\st}}(\blab_{\so};\blac_{\so}|\bmub_{\st};\bmuc_{\st}).
 \end{multline}
Here all the Bethe parameters are generic complex numbers and the sum is taken over the partitions of the sets $\blac$, $\blab$, $\bmuc$, and $\bmub$
 \be{part-1}
 \begin{array}{ll}
 \blac\Rightarrow\{\blac_{\so},\;\blac_{\st}\}, &\qquad  \bmuc\Rightarrow\{\bmuc_{\so},\;\bmuc_{\st}\},\\
 \blab\Rightarrow\{\blab_{\so},\;\blab_{\st}\}, &\qquad  \bmub\Rightarrow\{\bmub_{\so},\;\bmub_{\st}\} .
 \end{array}
 \ee
The partitions are independent except $\#\blac_{\so}=\#\blab_{\so}=a_{\so}$ and $\#\bmuc_{\so}=\#\bmub_{\so}=b_{\so}$. Accordingly one has
$\#\blac_{\st}=\#\blab_{\st}=a_{\st}=a-a_{\so}$ and $\#\bmuc_{\st}=\#\bmub_{\st}=b_{\st}=b-b_{\so}$.

The rational functions  $Z_{a_{\st},b_{\so}}$  and $Z_{a_{\so},b_{\st}}$ are the so-called highest coefficients.
They are equal to a partition function of $15$-vertex model with special boundary conditions \cite{Res86}. The reader can find their
explicit representations in \cite{Whe12,BelPRS12a}. We do not use these explicit formulas in the present paper  except
$Z_{0,0}(\emptyset;\emptyset|\emptyset;\emptyset)=1$. This condition is needed to satisfy the normalization $\mathcal{S}_{0,0}=\langle0|0\rangle=1$
(see \eqref{vac-def}).
Note also that the subscripts  of the highest coefficient are equal to the cardinalities of the subsets to the left and to the right of the
vertical line.

Form factors of the monodromy matrix entries are defined as
 \be{SP-deFF-gen}
 \mathcal{F}_{a,b}^{(i,j)}(z)\equiv\mathcal{F}_{a,b}^{(i,j)}(z|\blac,\bmuc;\blab,\bmub)=
 \mathbb{C}_{a',b'}(\blac;\bmuc)T_{ij}(z)\mathbb{B}_{a,b}(\blab;\bmub),
 \ee
where both $\mathbb{C}^{a',b'}(\blac;\bmuc)$ and $\mathbb{B}^{a,b}(\blab;\bmub)$ are on-shell
Bethe vectors, and
\be{apabpb}
\begin{array}{l}
a'=a+\delta_{i1}-\delta_{j1},\\
b'=b+\delta_{j3}-\delta_{i3}.
\end{array}
\ee
The parameter $z$ is an arbitrary complex  number. We call it  the external parameter.

It was proved in \cite{PakRS15a} that if $\{\blac,\bmuc\}\ne \{\blab,\bmub\}$, then the combination
\be{Univ-FF}
\mathfrak{F}_{a,b}^{(i,j)}(\blac,\bmuc;\blab,\bmub)=
\frac{\mathcal{F}_{a,b}^{(i,j)}(z|\blac,\bmuc;\blab,\bmub)}
{\tau(z|\blac,\bmuc)-\tau(z|\blab,\bmub)}
\ee
does not depend on $z$. We call $\mathfrak{F}_{a,b}^{(i,j)}(\blac,\bmuc;\blab,\bmub)$ the {\it universal form
factor} of the operator $T_{ij}(z)$. If $\blac\cap\blab=\emptyset$ and $\bmuc\cap\bmub=\emptyset$, then
the universal form factor is determined by the $R$-matrix only. It does not depend on a specific model,
in particular, on the functions $r_1(z)$ and $r_3(z)$. Determinant representations for $\mathfrak{F}_{a,b}^{(i,j)}$
were obtained in \cite{BelPRS12b,BelPRS13a,PakRS14b,PakRS14c,PakRS15a}.

Due to the invariance of the
$R$-matrix under transposition with respect to both spaces, the mapping
\be{def-psi}
\psi\,:
T_{ij}(u) \quad\mapsto\quad T_{ji}(u)
\ee
defines an antimorphism of the algebra \eqref{RTT}. The mapping \eqref{def-psi} acts in the algebra  \eqref{RTT}, therefore
expectation values of the operators $T_{ij}(z)$ are invariant under the action of $\psi$. In particular,
 \be{psi-FF1}
\psi\left( \mathcal{F}_{a,b}^{(i,j)}(z|\blac,\bmuc;\blab,\bmub)\right) = \mathcal{F}_{a,b}^{(i,j)}(z|\blac,\bmuc;\blab,\bmub).
\ee
On the other hand, we have
 \be{psi-FF2}
\psi\left( \mathbb{C}_{a',b'}(\blac;\bmuc)T_{ij}(z)\mathbb{B}_{a,b}(\blab;\bmub)\right)=
\mathbb{C}_{a,b}(\blab;\bmub)T_{ji}(z)\mathbb{B}_{a',b'}(\blac;\bmuc),
 \ee
and we recognize  the form factor of the operator $T_{ji}$ in the r.h.s. Thus, we obtain simple relations between different form factors:
\be{UF-UF}
\begin{aligned}
\mathcal{F}_{a,b}^{(i,j)}(z|\blac,\bmuc;\blab,\bmub)&=\mathcal{F}_{a',b'}^{(j,i)}(z|\blab,\bmub;\blac,\bmuc),\\
\mathfrak{F}_{a,b}^{(i,j)}(\blac,\bmuc;\blab,\bmub)&=
-\mathfrak{F}_{a',b'}^{(j,i)}(\blab,\bmub;\blac,\bmuc).
\end{aligned}
\ee

In the TCBG model the antimorphism $\psi$ implies the following mapping of the Bose fields \cite{Sla15}:
\be{psi-Psi}
\psi\bigl(\Psi_i(x)\bigr)=-\Psi^\dagger_i(L-x), \qquad \psi\bigl(\Psi^\dagger_i(x)\bigr)=-\Psi_i(L-x).
\ee
Due to \eqref{psi-Psi} we can relate form factors of the fields $\Psi_i(L-x)$ and $\Psi^\dagger_i(x)$
 \be{psi-FFPsi}
\psi\left( \mathbb{C}_{a',b'}(\blac;\bmuc)\Psi_i(L-x)\mathbb{B}_{a,b}(\blab;\bmub)\right)=
-\mathbb{C}_{a,b}(\blab;\bmub)\Psi^\dagger_i(x)\mathbb{B}_{a',b'}(\blac;\bmuc).
 \ee
Thus, it is enough to calculate the form factors of the fields $\Psi_i(x)$. The form factors of $\Psi^\dagger_i(x)$
can be obtained from the latter via \eqref{psi-FFPsi}.

\subsection{Composite model\label{SS-CM}}

In the   composite model the  total monodromy matrix $T(u)$  is
presented as a usual matrix product of the partial monodromy matrices $T^{(2)}(u)$ and $T^{(1)}(u)$:
\be{T-TT}
T(u)=T^{(2)}(u)T^{(1)}(u).
\ee
The matrix elements of $T(u)$ are operators in the space of states $\mathcal{H}$ that corresponds to an interval $[0,L]$.
The matrix elements of the partial monodromy matrices $T^{(1)}(u)$ and $T^{(2)}(u)$  act in the
spaces $\mathcal{H}^{(1)}$ and $\mathcal{H}^{(2)}$  corresponding
to the intervals $[0,x]$ and $[x,L]$ respectively. Here $x$ is an intermediate point of the interval $[0,L]$.
The total space of states $\mathcal{H}$ is a tensor product of the partial spaces of states $\mathcal{H}^{(1)}\otimes \mathcal{H}^{(2)}$.
The operators $T_{ij}^{(2)}(u)$ and $ T_{kl}^{(1)}(v)$ commute one with each other, as they act in different spaces.

Every $T^{(l)}(u)$ satisfies $RTT$-relation \eqref{RTT} and has its own vacuum state $|0\rangle^{(l)}$ and a dual state
$\langle0|^{(l)}$. Hereby
$|0\rangle= |0\rangle^{(1)}\otimes|0\rangle^{(2)}$ and $\langle0|=\langle0|^{(1)}\otimes \langle0|^{(2)}$.

The properties of the partial vacuum vectors  are similar to their total analogs, in particular,
\be{eigen}
T_{kk}^{(l)}(u)|0\rangle^{(l)}= r_{k}^{(l)}(u)|0\rangle^{(l)},\qquad
\langle0|^{(l)}T_{kk}^{(l)}(u)= r_{k}^{(l)}(u)\langle0|^{(l)}, \qquad l=1,2,
\ee
where $r_{k}^{(l)}(u)$ are some complex valued functions for $k=1,3$ and  $r_{2}^{(l)}(u)=1$. In the TCBG model
we have $r_{1}^{(l)}(u)=1$, $r_{3}^{(1)}(u)=e^{iux}$, and $r_{3}^{(2)}(u)=e^{iu(L-x)}$. Evidently
\be{lr}
r_{k}(u)=r_{k}^{(1)}(u)r_{k}^{(2)}(u).
\ee

Below we  express form factors in terms of $r_{k}^{(1)}(u)$, therefore we introduce a special notation for
these functions
\be{ell}
r_{k}^{(1)}(u)=\ell_k(u),\qquad \text{and hence,}\qquad r_{k}^{(2)}(u)=\frac{r_k(u)}{\ell_k(u)}, \qquad k=1,3.
\ee
Thus, in the TCBG model $\ell_1(u)=1$ and $\ell_3(u)=e^{iux}$, however, up to a certain point we continue to use the
notation $\ell_k(u)$ without its specification.

We extend convention \eqref{SH-prod} to the products of the functions $r_{k}^{(l)}(u)$ and $\ell_k(u)$. Namely, whenever these
functions depend on a set of variables  this means the product over the
corresponding set.

Finally, we recall the formulas for total on-shell (dual) Bethe vector in terms of partial (dual) Bethe vectors. They have the form \cite{PakRS15b,TarVar93}
\be{BV-BV-1}
\mathbb{B}_{a,b}(\bla;\bmu)=\sum \frac{\ell_{3}(\bmu_{\st})}{\ell_{1}(\bla_{\so})} f(\bla_{\so},\bla_{\st})f(\bmu_{\st},\bmu_{\so})f(\bmu_{\so},\bla_{\so})\;
\mathbb{B}_{a_{\so},b_{\so}}^{(1)}(\bla_{\so};\bmu_{\so}) \mathbb{B}_{a_{\st},b_{\st}}^{(2)}(\bla_{\st};\bmu_{\st}),
\ee
and
\be{CV-CV-eig}
\mathbb{C}_{a,b}(\bla;\bmu)=\sum \frac{\ell_{1}(\bla_{\st})} {\ell_{3}(\bmu_{\so})}f(\bla_{\so},\bla_{\st})f(\bmu_{\st},\bmu_{\so})f(\bmu_{\so},\bla_{\so})\;
\mathbb{C}_{a_{\so},b_{\so}}^{(1)}(\bla_{\so};\bmu_{\so}) \mathbb{C}_{a_{\st},b_{\st}}^{(2)}(\bla_{\st};\bmu_{\st}).
\ee
In \eqref{BV-BV-1} $\mathbb{B}_{a,b}$ is an on-shell Bethe vector of the total monodromy matrix $T(u)$, while $\mathbb{B}_{a_{\so},b_{\so}}^{(l)}$ are
Bethe vectors of the partial monodromy matrices $T^{(l)}(u)$ (partial Bethe vectors). Similarly equation \eqref{CV-CV-eig} expresses a dual total
on-shell Bethe vectors in terms of  partial dual Bethe vectors. In both formulas the sums are taken over all possible partitions
$\bla\Rightarrow\{\bla_{\so},\bla_{\st}\}$ and $\bmu\Rightarrow\{\bmu_{\so},\bmu_{\st}\}$. The cardinalities of the subsets are shown
by the subscripts of (dual) partial Bethe vectors and they run through all possible values.

\subsection{Total and partial zero modes}

The most principal difference between the TCBG model and the models considered in \cite{PakRS15c} appears in the definition of the monodromy
matrix zero modes. It was assumed in \cite{PakRS15c} that the monodromy matrix $T(u)$ goes to the identity operator at $|u|\to\infty$. In
the TCBG model this is true only for the left-upper $2\times 2$ block of $T(u)$. The properties of the zero modes in the TCBG model and
their relations to the Bose fields were found in \cite{Sla15}. In this section we list several results necessary for further applications.

We consider the zero modes of the total monodromy matrix $T(u)$ and partial zero modes of the partial monodromy matrices $T^{(l)}(u)$
(mainly for $l=1$). For $i,j=1,2$ the matrix elements $T_{ij}(u)$ and $T^{(l)}_{ij}(u)$ have the following asymptotic expansions:
\be{zero-modes}
\begin{aligned}
&T_{ij}(u)=\delta_{ij}+ \sum_{n=0}^\infty T_{ij}[n]\,\left(\frac cu\right)^{n+1},\\
&T^{(l)}_{ij}(u)=\delta_{ij}+ \sum_{n=0}^\infty T^{(l)}_{ij}[n]\,\left(\frac cu\right)^{n+1},\qquad l=1,2,
\end{aligned}
\qquad |u|\to\infty.
\ee
Accordingly the total and partial zero modes are defined as
\be{0-testij}
\begin{aligned}
&T_{ij}[0]=\lim_{|u|\to\infty}\frac uc(T_{ij}(u)-\delta_{ij}),\\
&T^{(l)}_{ij}[0]=\lim_{|u|\to\infty}\frac uc(T^{(l)}_{ij}(u)-\delta_{ij}), \qquad l=1,2,
\end{aligned}
\qquad i,j=1,2.
\ee
It is easy to see that $T_{ij}[0]=T^{(1)}_{ij}[0]+T^{(2)}_{ij}[0]$.

The partial zero modes $T^{(1)}_{ij}[0]$ ($i,j=1,2$) have the following explicit representation in terms of the Bose fields
$\Psi_i$ and $\Psi_i^\dagger$:
\be{0Tij-x}
T^{(1)}_{ij}[0]=-\int_0^x \Psi_i^\dagger(y)\Psi_j(y)\,dy, \qquad i,j=1,2.
\ee
Thus, computing form factors of these zero modes and taking the derivative over $x$ we obtain form factors of local operators
$\Psi_i^\dagger(x)\Psi_j(x)$.

The action of the total and partial zero modes $T_{ii}[0]$ and $T^{(1)}_{ii}[0]$  ($i=1,2$) onto the corresponding Bethe vectors has the following form
\be{actBVd-1}
\begin{aligned}
T_{11}[0]\mathbb{B}_{a,b}(\bla;\bmu)&=-a\mathbb{B}_{a,b}(\bla;\bmu), \\
T_{22}[0]\mathbb{B}_{a,b}(\bla;\bmu)&=(a-b)\mathbb{B}_{a,b}(\bla;\bmu),
\end{aligned}
\qquad
\begin{aligned}
T^{(1)}_{11}[0]\mathbb{B}^{(1)}_{a,b}(\bla;\bmu)&=-a\mathbb{B}^{(1)}_{a,b}(\bla;\bmu), \\
T^{(1)}_{22}[0]\mathbb{B}^{(1)}_{a,b}(\bla;\bmu)&=(a-b)\mathbb{B}^{(1)}_{a,b}(\bla;\bmu).
\end{aligned}
\ee
In these formulas $\mathbb{B}_{a,b}(\bla;\bmu)$ and $\mathbb{B}^{(1)}_{a,b}(\bla;\bmu)$ respectively are generic total and partial Bethe
vectors, i.e. their Bethe parameters are generic complex numbers.

The action of the zero modes $T_{12}[0]$ and $T_{21}[0]$ (and their partial analogs) is
\be{actBVa-1}
\begin{aligned}
T_{12}[0]\mathbb{B}_{a,b}(\bla;\bmu)&=\lim_{|w|\to\infty} \tfrac wc\;\mathbb{B}_{a+1,b}(\{w,\bla\};\bmu), \\
\mathbb{C}_{a,b}(\bla;\bmu)T_{21}[0]&=\lim_{|w|\to\infty} \tfrac wc\;\mathbb{C}_{a+1,b}(\{w,\bla\};\bmu),
\end{aligned}
\qquad
\begin{aligned}
T^{(1)}_{12}[0]\mathbb{B}^{(1)}_{a,b}(\bla;\bmu)&=\lim_{|w|\to\infty} \tfrac wc\;\mathbb{B}^{(1)}_{a+1,b}(\{w,\bla\};\bmu), \\
\mathbb{C}^{(1)}_{a,b}(\bla;\bmu)T^{(1)}_{21}[0]&=\lim_{|w|\to\infty} \tfrac wc\;\mathbb{C}^{(1)}_{a+1,b}(\{w,\bla\};\bmu).
\end{aligned}
\ee
Here also the (dual) Bethe vectors are generic. It is important to note, however, that if the (dual) Bethe vectors are on-shell, then the resulting
vectors also are on-shell, because the Bethe equations have infinite roots of $u$-type.

Finally we need singular properties of total on-shell (dual) Bethe vectors
\be{hwC}
\mathbb{C}_{a,b}(\bla;\bmu)\,T_{12}[0]=0,\qquad T_{21}[0]\,\mathbb{B}_{a,b}(\bla;\bmu)=0.
\ee
Here $\mathbb{C}_{a,b}(\bla;\bmu)$ and $\mathbb{B}_{a,b}(\bla;\bmu)$ are on-shell Bethe vectors.
This property was found in \cite{MuhTV06} for $GL(N)$-invariant models. In the $GL(3)$ case it  follows from
the explicit formulas of the action of the operators $T_{ij}(z)$ onto Bethe vectors \cite{BelPRS12c}.

The definitions of the (total and partial) zero modes of the operators $T_{i3}$ and $T_{3i}$ are different from \eqref{zero-modes}. The operator $T_{33}(u)$ has the following expansion
\be{zero-modes33}
\begin{aligned}
&T_{33}(u)=e^{iLu}+ e^{iLu}\sum_{n=0}^\infty T_{33}[n]\,\left(\frac cu\right)^{n+1},\\
&T^{(1)}_{33}(u)=e^{ixu}+ e^{ixu}\sum_{n=0}^\infty T^{(1)}_{33}[n]\,\left(\frac cu\right)^{n+1}.
\end{aligned}
\ee
Respectively the total and partial zero modes are defined as
\be{0-testij33}
\begin{aligned}
&T_{33}[0]=\lim_{|u|\to\infty}\frac uc(e^{-iLu}T_{33}(u)-1),\\
&T^{(1)}_{33}[0]=\lim_{|u|\to\infty}\frac uc(e^{-ixu}T^{(1)}_{33}(u)-1).
\end{aligned}
\ee

Expansions \eqref{zero-modes}, \eqref{0-testij33} imply the following expansion of the functions $\ell_k(u)$:
\be{zero-modesl3}
\begin{aligned}
&\ell_1(u)=1+ \ell_1[0]\,\frac cu +o(u^{-1}),\\
&\ell_3(u)=e^{ixu}\Bigl(1+ \ell_3[0]\,\frac cu +o(u^{-1})\Bigr),
\end{aligned}
\qquad |u|\to\infty .
\ee
Since $\ell_1(u)=1$ and  $\ell_3(u)=e^{ixu}$, we conclude that $\ell_1[0]=\ell_3[0]=0$.

It turns out that in the TCBG model $T_{33}[0]=-T_{11}[0]-T_{22}[0]$ (and similarly for the partial zero modes), therefore
below we do not consider these zero modes.

Asymptotic expansions of the operators  $T_{i3}$ and $T_{3i}$ with $i=1,2$ are more sophisticated. We give them for the
partial zero modes of the operators $T^{(1)}_{i3}(u)$ and $T^{(1)}_{3j}(u)$
\begin{equation}\label{Ti3-3j}
\begin{aligned}
&T^{(1)}_{i3}(u)=-\frac{\sqrt{\cc}}{u}\left(e^{iux}\Psi_{i}^\dagger(x)-\Psi_{i}^\dagger(0)\right)+O(u^{-2}),\qquad i=1,2,\\
&T^{(1)}_{3j}(u)=-\frac{\sqrt{\cc}}{u}\left(\Psi_{j}(x)-e^{iux}\Psi_{j}(0)\right)+O(u^{-2}),\qquad j=1,2.
\end{aligned}
 \end{equation}
For the total zero modes one should replace $x$ by $L$ everywhere in these formulas.
We see that the asymptotic behavior $|u|\to\infty$ leads to two types of zero modes,  corresponding to the two boundaries of the interval $[0,x]$:
the left (resp. right) boundary corresponds to the left partial zero modes $\ZMLp{ij}$ (resp. the right ones $\ZMRp{ij}$).
For our goal we  need
only the right partial zero modes, which are defined as follows:
\be{0Ti3-x}
\begin{aligned}
&\ZMRp{i3}=\lim_{u\to-i\infty}e^{-iux}\frac uc \;T^{(1)}_{i3}(u), \\
&\ZMRp{3i}=\lim_{u\to+i\infty} \frac uc \;T^{(1)}_{3i}(u),
\end{aligned}
\qquad i=1,2,
\ee
and hence,
\be{0Ti3-xpsi}
\ZMRp{i3}=\frac1{i\sqrt{\cc}}\;\Psi^\dagger_i(x),\qquad \ZMRp{3i}= \frac1{i\sqrt{\cc}}\;\Psi_i(x), \qquad i=1,2.
\ee
Thus, calculating the form factors of these zero modes leads to the evaluation of the  form factors of the local fields
$\Psi_i(x)$ and $\Psi^\dagger_i(x)$.

Below we will need the actions of the right partial zero modes $\ZMRp{23}$ and $\ZMRp{32}$ respectively onto usual and dual
partial Bethe vectors:
\be{actBVd-2}
\begin{aligned}
&\ZMRp{23}\mathbb{B}^{(1)}_{a,b}(\bla;\bmu)=\lim_{w\to-i\infty} e^{-iwx} \tfrac wc\;\mathbb{B}^{(1)}_{a,b+1}(\bla;\{w,\bmu\}),\\
&\mathbb{C}^{(1)}_{a,b}(\bla;\bmu)\ZMRp{32}=\lim_{w\to+i\infty}\tfrac wc\;\mathbb{C}^{(1)}_{a,b+1}(\bla;\{w,\bmu\}).
\end{aligned}
\ee
Here both Bethe vectors are generic.

\section{Main results\label{S-MR}}

In this section we give a list of formulas for the form factors of local operators of the TCBG model in terms of the universal
form factors \eqref{Univ-FF}. The reader can find determinant representations for the universal
form factors in \cite{BelPRS12b,BelPRS13a,PakRS14b,PakRS14c,PakRS15a}.

For given on-shell vectors $\mathbb{C}_{a',b'}(\blac;\bmuc)$ and  $\mathbb{B}_{a,b}(\blac;\bmuc)$ define an excitation momentum as
\be{Ex-Mom}
\mathcal{P}(\bmub,\bmuc)=\sum_{i=1}^b\mub_i-\sum_{i=1}^{b'}\muc_i.
\ee

\begin{thm}\label{Theor-FF-Qij}
Let $\mathbb{C}_{a',b'}(\blac;\bmuc)$ and $\mathbb{B}_{a,b}(\blac;\bmuc)$ be on-shell Bethe vectors such that $\{\blac,\bmuc\}
\ne \{\blab,\bmub\}$. Then the form factors of the operators $\Psi^\dagger_i(x)\Psi_j(x)$ ($i,j=1,2$) have the following representation:
\begin{equation}\label{FF-pipj-res}
\mathbb{C}_{a',b}(\blac;\bmuc)\Psi_i^\dagger(x)\Psi_j(x)\mathbb{B}_{a,b}(\blab;\bmub)=-i\mathcal{P}(\bmub,\bmuc)
\;e^{ix\mathcal{P}(\bmub,\bmuc)}
\mathfrak{F}_{a,b}^{(i,j)}(\blac,\bmuc;\blab,\bmub),
\end{equation}
where $\mathfrak{F}_{a,b}^{(i,j)}(\blac,\bmuc;\blab,\bmub)$ is the universal form factor of the matrix element $T_{ij}(z)$
and $a'=a+j-i$.
\end{thm}

Theorem~\ref{Theor-FF-Qij}  is a direct corollary of determinant representations for the partial zero modes
obtained in \cite{PakRS15c}
\begin{equation}\label{FF-Tee-1}
\mathbb{C}_{a',b'}(\blac;\bmuc)\;T^{(1)}_{ij}[0]\;\mathbb{B}_{a,b}(\blab;\bmub)=
\left(\frac{\ell_1(\blac)\ell_3(\bmub)}{\ell_1(\blab)\ell_3(\bmuc)}-1\right)\mathfrak{F}_{a,b}^{(i,j)}(\blac,\bmuc;\blab,\bmub),
\end{equation}
where $a'=a+\delta_{i1}-\delta_{j1}$, $b'=b+\delta_{j3}-\delta_{i3}$. We have seen that for the matrix elements $T^{(1)}_{ij}(u)$ with
$i,j=1,2$ the actions of the zero modes onto Bethe vectors in the TCBG model  are the same as in the models considered in \cite{PakRS15c}. Therefore the form factors
of the partial zero modes also are the same.  One should
only specify $\ell_1(u)=1$, $\ell_3(v)=e^{ixv}$ in \eqref{FF-Tee-1} and use \eqref{0Tij-x}.

Let $\bar\kappa=\{\kappa_1,\kappa_2,\kappa_3\}$. Consider the following deformation of the Bethe equations \eqref{AEigenS-1}
\be{ATEigenS-1}
1=\frac{\kappa_2}{\kappa_1}\frac{f(\bla_{\so},\bla_{\st})}{f(\bla_{\st},\bla_{\so})}f(\bmu,\bla_{\so}),\qquad
r_3(\bmu_{\so})=\frac{\kappa_2}{\kappa_3}\frac{f(\bmu_{\st},\bmu_{\so})}{f(\bmu_{\so},\bmu_{\st})}f(\bmu_{\so},\bla),
\ee
where $\#\bla_{\so}=\#\bmu_{\so}=1$. This system is called twisted Bethe equations. It determines
the roots $v_i$ and $u_i$ as implicit functions of the parameters $\bar\kappa$: $v_i=v_i(\bar\kappa)$ and $u_i=u_i(\bar\kappa)$.

\begin{thm}\label{Theor-FF-Qii}
Let $\{\blac,\bmuc\}= \{\blab,\bmub\}=\{\bla,\bmu\}$. Then the form factors of the operators $\Psi^\dagger_j(x)\Psi_j(x)$ ($j=1,2$) have the following representation:
\begin{equation}\label{FF-pipi-res}
\mathbb{C}_{a,b}(\bla;\bmu)\;\Psi_j^\dagger(x)\Psi_j(x)\;\mathbb{B}_{a,b}(\bla;\bmu)=
i\sum_{k=1}^b\frac{dv_k(\bar\kappa)}{d\kappa_j}
\Bigr|_{\bar\kappa=1}\cdot\|\mathbb{B}_{a,b}(\bla;\bmu)\|^2, \qquad j=1,2,
\end{equation}
\begin{equation}\label{FF-pipi-sum}
\mathbb{C}_{a,b}(\bla;\bmu)\;\bigl(\Psi_1^\dagger(x)\Psi_1(x)+\Psi_2^\dagger(x)\Psi_2(x)\bigr)\;\mathbb{B}_{a,b}(\bla;\bmu)=
\frac bL\;\|\mathbb{B}_{a,b}(\bla;\bmu)\|^2,
\end{equation}
where $\bmu(\bar\kappa)$ is a deformation of $\bmu$, such that the set $\bmu(\bar\kappa)$ satisfies twisted Bethe equations \eqref{ATEigenS-1} and
$\bmu(\bar\kappa)=\bmu$ at $\bar\kappa=\{1,1,1\}$.
\end{thm}

Equation \eqref{FF-pipi-res} of theorem~\ref{Theor-FF-Qii} also directly follows from the corresponding representation for the expectation value of the partial zero modes obtained in \cite{PakRS15c}
\begin{equation}\label{FF-Tee-2}
\mathbb{C}_{a,b}(\bla;\bmu)\;T^{(1)}_{ii}[0]\;\mathbb{B}_{a,b}(\bla;\bmu)=
\left(\delta_{i,1}\ell_1[0]+\delta_{i,3}\ell_3[0]+\frac{d}{d\kappa_i}
\log\frac{\ell_1\bigl(\bla(\bar\kappa)\bigr)}{\ell_3\bigl(\bmu(\bar\kappa)\bigr)}
\Bigr|_{\bar\kappa=1}\right)\;\|\mathbb{B}_{a,b}(\bla;\bmu)\|^2.
\end{equation}
Setting here $\ell_1[0]=\ell_3[0]=0$, $\ell_1(u)=1$, and $\ell_3(v)=e^{ixv}$ we immediately arrive at \eqref{FF-pipi-res}. Furthermore,
for a special choice $\kappa_1=\kappa_2=\kappa$, $\kappa_3=1$ equation \eqref{FF-Tee-2} takes the following form:
\begin{equation}\label{FF-Tee-2sum}
\mathbb{C}_{a,b}(\bla;\bmu)\bigl(T^{(1)}_{11}[0]+T^{(1)}_{22}[0]\bigr)\mathbb{B}_{a,b}(\bla;\bmu)=
\left(\ell_1[0]+\frac{d}{d\kappa}
\log\frac{\ell_1\bigl(\bla(\kappa)\bigr)}{\ell_3\bigl(\bmu(\kappa)\bigr)}
\Bigr|_{\kappa=1}\right)\;\|\mathbb{B}_{a,b}(\bla;\bmu)\|^2.
\end{equation}
It is easy to see that for this special choice of $\bar\kappa$
 the system \eqref{ATEigenS-1} has a very simple solution in terms of non-deformed Bethe parameters
\be{deform}
u_k(\bar\kappa)=u_k-\frac iL\log\kappa, \qquad v_k(\bar\kappa)=v_k-\frac iL\log\kappa,
\ee
where $\bla$ and $\bmu$ are solutions of the standard Bethe equations \eqref{AEigenS-1}. Substituting  \eqref{deform} into
\eqref{FF-Tee-2sum} we obtain \eqref{FF-pipi-sum}.

\begin{thm}\label{Theor-FF-field}
The form factors of the Bose fields  $\Psi_k(x)$ and  $\Psi^\dagger_k(x)$  ($k=1,2$)  have the following representation:
\be{FF-psik-res}
\mathbb{C}_{a-2+k,b-1}(\blac;\bmuc)\Psi_k(x)\mathbb{B}_{a,b}(\blab;\bmub)
=i\sqrt{\cc}\;e^{ix\mathcal{P}(\bmub,\bmuc)}\mathfrak{F}_{a,b}^{(3,k)}(\blac,\bmuc;\blab,\bmub),
\ee
\be{FF-psik-res+}
\mathbb{C}_{a+2-k,b+1}(\blac;\bmuc)\Psi^\dagger_k(x)\mathbb{B}_{a,b}(\blab;\bmub)
=i\sqrt{\cc}\;e^{ix\mathcal{P}(\bmub,\bmuc)}\mathfrak{F}_{a,b}^{(k,3)}(\blac,\bmuc;\blab,\bmub),
\ee
where $\mathfrak{F}_{a,b}^{(3,k)}(\blac,\bmuc;\blab,\bmub)$ and $\mathfrak{F}_{a,b}^{(k,3)}(\blac,\bmuc;\blab,\bmub)$ are respectively the universal
form factors of the matrix elements $T_{3k}(z)$ and $T_{k3}(z)$.
\end{thm}

The statement of this theorem cannot be obtained directly from the results of \cite{PakRS15c}. Here we show that if
equation \eqref{FF-psik-res} holds for $k=2$, then it is valid for $k=1$, and then \eqref{FF-psik-res+} is also valid for
$k=1,2$. The proof of equation \eqref{FF-psik-res} for $k=2$ will be given in the next section.

Let us denote form factors of the partial zero modes $\ZMRp{ij}$ as
\be{Not-ffij}
{\sf M}_{a,b}^{i,j}(x)\equiv{\sf M}_{a,b}^{i,j}(x|\blac,\bmuc;\blab,\bmub)=\mathbb{C}_{a',b'}(\blac;\bmuc)\ZMRp{ij}\mathbb{B}_{a,b}(\blab;\bmub).
\ee
Recall that here $a'=a+\delta_{i1}-\delta_{j1}$, $b'=b+\delta_{j3}-\delta_{i3}$.
It follows from  \eqref{0Ti3-xpsi} that the form factors of fields $\Psi_k(x)$ can be obtained from the form factors
${\sf M}_{a,b}^{3,k}$, $k=1,2$. Let us show that ${\sf M}_{a,b}^{3,1}$ and ${\sf M}_{a,b}^{3,2}$ are related to each other by a simple
limiting procedure.

\begin{prop}
\be{M32M31}
\lim_{w\to+i\infty}\frac wc {\sf M}_{a,b}^{3,2}(x|\{\blac,w\},\bmuc;\blab,\bmub)={\sf M}_{a,b}^{3,1}(x|\blac,\bmuc;\blab,\bmub).
\ee
\end{prop}

{\sl Proof.} We have from $RTT$-relation \eqref{RTT}
\be{Com-2132}
[T^{(1)}_{21}(u),T^{(1)}_{32}(v)]=g(u,v)\bigl(T^{(1)}_{31}(v)T^{(1)}_{22}(u)-T^{(1)}_{31}(u)T^{(1)}_{22}(v)\bigr).
\ee
Multiplying this equation by $c^2/(uv)$ and sending $u,v\to+i\infty$ we obtain
\be{Com-2132-zero}
[T^{(1)}_{21}[0],\ZMRp{32}]=\ZMRp{31},
\ee
and thus,
\be{Com-2132-part}
[T_{21}[0],\ZMRp{32}]=\ZMRp{31},
\ee
because $T_{21}[0]=T^{(1)}_{21}[0]+T^{(2)}_{21}[0]$ and $T^{(2)}_{21}[0]$ commutes with $\ZMRp{32}$.
Due to \eqref{actBVa-1} we have
\be{M32lim}
\lim_{w\to+i\infty}\frac wc {\sf M}_{a,b}^{3,2}(x|\{\blac,w\},\bmuc;\blab,\bmub)=
\mathbb{C}_{a-1,b-1}(\blac;\bmuc) T_{21}[0]\ZMRp{32}\mathbb{B}_{a,b}(\blab;\bmub).
\ee
Since the action of $T_{21}[0]$ on the on-shell vector $\mathbb{B}_{a,b}(\blab;\bmub)$ gives zero (see \eqref{hwC}),
we can replace in \eqref{M32lim} the product $T_{21}[0]\ZMRp{32}$ by the commutator $[T_{21}[0],\ZMRp{32}]$.
The last one is equal to $\ZMRp{31}$ due to \eqref{Com-2132-part}. We arrive at
\be{M32lim-take}
\lim_{w\to+i\infty}\frac wc {\sf M}_{a,b}^{3,2}(x|\{\blac,w\},\bmuc;\blab,\bmub)=
\mathbb{C}_{a-1,b-1}(\blac;\bmuc) \ZMRp{31}\mathbb{B}_{a,b}(\blab;\bmub),
\ee
which ends the proof.

It remains to take this limit in \eqref{FF-psik-res} for $k=2$. Using (see \cite{PakRS15a})
\be{F32F31}
\lim_{w\to+i\infty}\frac wc \mathfrak{F}_{a,b}^{3,2}(\{\blac,w\},\bmuc|\blab,\bmub)=\mathfrak{F}_{a,b}^{3,1}(\blac,\bmuc|\blab,\bmub),
\ee
and $\ell_1(w)=1$ we conclude that
if representation \eqref{FF-psik-res} holds for the form factor of $\Psi_2(x)$, then it holds for the form factor of $\Psi_1(x)$.

Now we should prove that \eqref{FF-psik-res} yields \eqref{FF-psik-res+}. For this we use the mapping \eqref{psi-Psi}.
Due to \eqref{0Ti3-xpsi} and \eqref{FF-psik-res} we have
\be{FF-psik-p1}
{\sf M}_{a,b}^{3,k}(x|\blacp,\bmucp;\blabp,\bmubp)
=\frac{\ell_3(\bmubp)}{\ell_3(\bmucp)}\mathfrak{F}_{a,b}^{(3,k)}(\blacp,\bmucp;\blabp,\bmubp).
\ee
Recall that here the sets $\{\blacp,\bmucp\}$ and $\{\blabp,\bmubp\}$ satisfy the Bethe equations.
Replacing in \eqref{FF-psik-p1} $x$ by $L-x$ we obtain
\be{FF-psik-p2}
{\sf M}_{a,b}^{3,k}(L-x|\blacp,\bmucp;\blabp,\bmubp)
=\frac{\ell_3(\bmucp)}{\ell_3(\bmubp)}\mathfrak{F}_{a,b}^{(3,k)}(\blacp,\bmucp;\blabp,\bmubp).
\ee
Here we have used
\be{have-used}
\frac{\ell_3(\bmubp)}{\ell_3(\bmucp)}\Bigr|_{x\to L-x}= \frac{r_3(\bmubp)}{r_3(\bmucp)}\frac{\ell_3(\bmucp)}{\ell_3(\bmubp)}
=\frac{\ell_3(\bmucp)}{\ell_3(\bmubp)},
\ee
because due to \eqref{prod-r} in the TCBG model $r_3(\bmubp)=r_3(\bmucp)=1$. Now we act on \eqref{FF-psik-p2} with the mapping
$\psi$ \eqref{psi-Psi}. The r.h.s. remains invariant, while in the l.h.s. we obtain due to \eqref{psi-FFPsi}
 \be{psi-FFPsi0}
\psi\left( {\sf M}_{a,b}^{3,k}(L-x|\blacp,\bmucp;\blabp,\bmubp)\right)=
-{\sf M}_{a,b}^{k,3}(x|\blabp,\bmubp;\blacp,\bmucp).
 \ee
Thus, we arrive at
\be{FF-prime}
{\sf M}_{a',b'}^{k,3}(x|\blabp,\bmubp;\blacp,\bmucp)=-\frac{\ell_3(\bmucp)}{\ell_3(\bmubp)}
\mathfrak{F}_{a,b}^{(3,k)}(\blacp,\bmucp;\blabp,\bmubp).
\ee
Now we simply rename the Bethe parameters
\be{rename-BP}
\blacp\to\blab,\qquad\blabp\to\blac, \qquad \bmucp\to\bmub,\qquad\bmubp\to\bmuc, \qquad \{a',b'\}\leftrightarrow \{a,b\},
\ee
and  use \eqref{UF-UF}. We obtain
\be{FF-not-prime}
{\sf M}_{a,b}^{k,3}(x|\blac,\bmuc;\blab,\bmub)=\frac{\ell_3(\bmub)}{\ell_3(\bmuc)}
\mathfrak{F}_{a,b}^{(k,3)}(\blac,\bmuc;\blab,\bmub).
\ee
This equation together with \eqref{0Ti3-xpsi} implies \eqref{FF-psik-res+}.

Thus, in order to prove the statements of theorem~\ref{Theor-FF-field} it is enough to prove \eqref{FF-psik-res} for $k=2$. This will be done in the
next section.

\section{Form factor of the field $\Psi_2(x)$\label{S-FFF2x}}

Due to \eqref{0Ti3-xpsi} the form factor of the field $\Psi_2(x)$ is related to the form factor ${\sf M}_{a,b}^{3,2}$ of the right
partial zero mode $\ZMRp{32}$.
The latter should be calculated straightforwardly using the action  formula
\eqref{actBVd-2} and the scalar product formula \eqref{Resh-SP}. We used similar way in \cite{PakRS15c} for the calculating
the form factors of the diagonal partial zero modes. That derivation was based on a summation identity that also plays a very important role
in the present case.

\subsection{Summation identities\label{S-SI}}

\begin{lemma}\label{Lem-FF1}
Let $\blacb$ and $\bmucb$ be four sets of generic complex numbers with cardinalities $\#\blacb=a$ and $\#\bmucb=b$, $a,b=0,1,\dots$.
For arbitrary partitions of these sets of the form \eqref{part-1} define a function $W$ as
 \begin{multline}\label{Resh-SP-eig2}
 W\begin{pmatrix}
  \blac_{\so},\blab_{\so},&\bmuc_{\so},\bmub_{\so}\\
    \blac_{\st},\blab_{\st},&\bmuc_{\st},\bmub_{\st}
  \end{pmatrix}=   f(\blac_{\st},\blac_{\so})  f(\blab_{\so},\blab_{\st})  f(\bmuc_{\so},\bmuc_{\st})
   f(\bmub_{\st},\bmub_{\so})f(\bmuc_{\st},\blac_{\st})f(\bmub_{\so},\blab_{\so})
  \num
 \times  \;Z_{a_{\st},b_{\so}}(\blac_{\st};\blab_{\st}|\bmuc_{\so};\bmub_{\so})
 Z_{a_{\so},b_{\st}}(\blab_{\so};\blac_{\so}|\bmub_{\st};\bmuc_{\st}),
 \end{multline}
where $Z_{a,b}$ are the highest coefficients (see \eqref{Resh-SP}). Then
 \begin{equation}\label{Wpart}
 \sum      W\begin{pmatrix}
  \blac_{\so},\blab_{\so},&\bmuc_{\so},\bmub_{\so}\\
    \blac_{\st},\blab_{\st},&\bmuc_{\st},\bmub_{\st}
  \end{pmatrix}=\delta_{0a}\delta_{0b},
 \end{equation}
where the sum is taken over all possible partitions $\blacb\Rightarrow\{\blacb_{\so},\;\blacb_{\st}\}$
and $\bmucb\Rightarrow\{\bmucb_{\so},\;\bmucb_{\st}\}$ with $\#\blac_{\so}=\#\blab_{\so}$ and $\#\bmuc_{\so}=\#\bmub_{\so}$.
\end{lemma}

Identity \eqref{Wpart} was proved in \cite{BelPRS12b}.
In order to compute the form factor ${\sf M}_{a,b}^{3,2}$ we need one more summation identity.

\begin{lemma}\label{Lem-FF32}
Let  $\blacb$ and $\bmucb$ be as in lemma~\ref{Lem-FF1} and $w$ be an arbitrary complex number.
Let also $\{\bmuc,w\}=\bxi$. Then
 \begin{equation}\label{W-T32}
  \lim_{|w|\to\infty}\frac wc \sum_{w\in\bxi_{\so}}
      W\begin{pmatrix}
  \blac_{\so},\blab_{\so},&\bxi_{\so},\bmub_{\so}\\
    \blac_{\st},\blab_{\st},&\bxi_{\st},\bmub_{\st}
  \end{pmatrix}=\mathfrak{F}_{a,b}^{(3,2)}(\blac,\bmuc;\blab,\bmub).
 \end{equation}
Here the sum is taken over partitions $\blacb\Rightarrow\{\blacb_{\so},\;\blacb_{\st}\}$, $\bmub\Rightarrow\{\bmub_{\so},\;\bmub_{\st}\}$,
and $\bxi\Rightarrow\{\bxi_{\so},\;\bxi_{\st}\}$ with $\#\blac_{\so}=\#\blab_{\so}$ and $\#\bxi_{\so}=\#\bmub_{\so}$, and we demand  that
$w\in\bxi_{\so}$. This restriction on the partitions is indicated explicitly by the subscript of the sum.
\end{lemma}

The proof of this lemma is given in appendix~\ref{A-PL}.

We would like to emphasise that the restriction $w\in\bxi_{\so}$ is of great importance. Without this restriction the sum in \eqref{W-T32}
vanishes due to lemma~\ref{Lem-FF1}. Therefore, in particular, equation \eqref{W-T32} implies
 \begin{equation}\label{W-T32-d}
  \lim_{|w|\to\infty}\frac wc \sum_{w\in\bxi_{\st}}
      W\begin{pmatrix}
  \blac_{\so},\blab_{\so},&\bxi_{\so},\bmub_{\so}\\
    \blac_{\st},\blab_{\st},&\bxi_{\st},\bmub_{\st}
  \end{pmatrix}=-\mathfrak{F}_{a,b}^{(3,2)}(\blac,\bmuc;\blab,\bmub),
 \end{equation}
where the sum is now taken with the restriction $w\in\bxi_{\st}$. Indeed, adding together the sums in \eqref{W-T32} and \eqref{W-T32-d} gives a sum
without any restriction, which is equal to zero.

\subsection{Derivation of a determinant representation\label{SS-PSI}}

Due to \eqref{actBVd-2} the action of $\ZMRp{32}$ on a total dual on-shell
Bethe vector is
\begin{multline}\label{act-BV-BV32}
\mathbb{C}_{a,b-1}(\blac;\bmuc) \ZMRp{32}=\lim_{w\to+i\infty} \frac{w}{c}\;\sum
\frac{\ell_{3}(\bmuc_{\st})}{\ell_{1}(\blac_{\so})} f(\blac_{\so},\blac_{\st})f(\bmuc_{\st},\bmuc_{\so})
f(\bmuc_{\so},\blac_{\so})\\
\times\mathbb{C}_{a_{\so},b_{\so}}^{(1)}(\blac_{\so};\{w,\bmuc_{\so}\}) \mathbb{C}_{a_{\st},b_{\st}}^{(2)}(\blac_{\st};\bmuc_{\st}).
\end{multline}
Let  $\{w,\bmuc\}=\bxi$. Then
\begin{multline}\label{Mat-el-32}
{\sf M}_{a,b}^{(3,2)}(x)=\lim_{w\to+i\infty} e^{iwx}\frac{w}{c}\;\sum
\frac{\ell_1(\blac_{\st})\ell_3(\bmub_{\st})}{\ell_1(\blab_{\so})\ell_3(\bxi_{\so})}\;
 f(\blac_{\so},\blac_{\st})f(\blab_{\so},\blab_{\st})f(\bxi_{\st},\bxi_{\so})f(\bmub_{\st},\bmub_{\so})
\\
\times f(\bmub_{\so},\blab_{\so})f(\bxi_{\so},\blac_{\so})\;
\mathbb{C}_{a_{\so},b_{\so}+1}^{(1)}(\blac_{\so};\bxi_{\so})\mathbb{B}_{a_{\so},b_{\so}+1}^{(1)}(\blab_{\so};\bmub_{\so}) \cdot \mathbb{C}_{a_{\st},b_{\st}}^{(2)}(\blac_{\st};\bxi_{\st})\mathbb{B}_{a_{\st},b_{\st}}^{(2)}(\blab_{\st};\bmub_{\st}).
\end{multline}
Here the sum in is taken over the partitions
$\blacb\Rightarrow\{\blacb_{\so},\blacb_{\st}\}$, $\bmub\Rightarrow\{\bmub_{\so},\bmub_{\st}\}$,
and $\bxi\Rightarrow\{\bxi_{\so},\bxi_{\st}\}$ with a restriction
$w\in\bxi_{\so}$.  We also have set  $\bxi_{\so}=\{w,\bmuc_{\so}\}$ and  $\bxi_{\st}=\bmuc_{\st}$.

Since $w\in\bxi_{\so}$ and $\ell_3(w)=e^{iwx}$, one goes from \eqref{act-BV-BV32} to \eqref{Mat-el-32} using
%
\be{repl-prod}
\ell^{-1}_3(\bmuc_{\so})=e^{iwx}\ell^{-1}_3(\bxi_{\so}).
\ee
%
In doing so, we also replaced the products $f(\bmuc_{\st},\bmuc_{\so})$ and $f(\bmuc_{\so},\blac_{\so})$ by
$f(\bxi_{\st},\bxi_{\so})$ and $f(\bxi_{\so},\blac_{\so})$ respectively. This is possible, because if the $f$ function  depends
on $w$, then it goes to $1$ in the limit $w\to+i\infty$.

We should substitute the explicit expression for the scalar products \eqref{Resh-SP} into \eqref{Mat-el-12}. It is clear that we
obtain new partitions of the subsets into subsubsets. Therefore, in order to avoid cumbersome roman numbers, we numerate
these subsubsets by the standard arabic numbers.

Using \eqref{Resh-SP} for the scalar product of $\mathbb{C}^{(1)}$ and $\mathbb{B}^{(1)}$ we should replace all the functions $r_k$ by $\ell_k$:
 \begin{multline}\label{Resh-SP-11}
\mathbb{C}^{(1)}_{a_{\so},b_{\so}+1}(\blac_{\so};\bxi_{\so})\mathbb{B}^{(1)}_{a_{\so},b_{\so}+1}(\blab_{\so};\bmub_{\so})=
\sum \ell_1(\blab_{1})\ell_1(\blac_{3}) \ell_3(\bxi_{3})\ell_3(\bmub_{1})  f(\blac_{1},\blac_{3})  f(\blab_{3},\blab_{1})     \num
 \times f(\bxi_{3},\bxi_{1})   f(\bmub_{1},\bmub_{3})
 \frac{f(\bxi_{1},\blac_{1})f(\bmub_{3},\blab_{3})}{f(\bxi_{\so},\blac_{\so})f(\bmub_{\so},\blab_{\so})}
  \;Z_{a_{3},b_{1}}(\blac_{3};\blab_{3}|\bxi_{1};\bmub_{1})
 Z_{a_{1},b_{3}}(\blab_{1};\blac_{1}|\bmub_{3};\bxi_{3}).
 \end{multline}
The summation is taken with respect to the partitions
\be{1-divis}
\blacb_{\so}\Rightarrow\{\blacb_1,\blacb_3\},\qquad \bmub_{\so}\Rightarrow\{\bmub_1,\bmub_3\},
\qquad \bxi_{\so}\Rightarrow\{\bxi_1,\bxi_3\}.
\ee
The cardinalities of the subsubsets are $a_n=\#\blacb_n$, $b_n=\#\bmub_n=\#\bxi_n$, $n=1,3$.

In the scalar product of $\mathbb{C}^{(2)}$ and
$\mathbb{B}^{(2)}$ we should replace the functions $r_k$ by $r_k/\ell_k$:
 \begin{multline}\label{Resh-SP-22}
\mathbb{C}^{(2)}_{a_{\st},b_{\st}}(\blac_{\st};\bxi_{\st})\mathbb{B}^{(2)}_{a_{\st},b_{\st}}(\blab_{\st};\bmub_{\st})
=\sum \frac{r_2(\blab_{2})r_2(\blac_{4}) r_4(\bxi_{4})r_4(\bmub_{2})}
{\ell_2(\blab_{2})\ell_2(\blac_{4}) \ell_4(\bxi_{4})\ell_4(\bmub_{2})}
  f(\blac_{2},\blac_{4})  f(\blab_{4},\blab_{2})     \num
 \times f(\bxi_{4},\bxi_{2})   f(\bmub_{2},\bmub_{4})
 \frac{f(\bxi_{2},\blac_{2})f(\bmub_{4},\blab_{4})}{f(\bxi_{\st},\blac_{\st})f(\bmub_{\st},\blab_{\st})}
  \;Z_{a_{4},b_{2}}(\blac_{4};\blab_{4}|\bxi_{2};\bmub_{2})
 Z_{a_{2},b_{4}}(\blab_{2};\blac_{2}|\bmub_{4};\bxi_{4}).
 \end{multline}
The summation is taken with respect to the partitions
\be{2-divis}
\blacb_{\st}\Rightarrow\{\blacb_2,\blacb_4\},\qquad \bmub_{\st}\Rightarrow\{\bmub_2,\bmub_4\},
\qquad \bxi_{\st}\Rightarrow\{\bxi_2,\bxi_4\}.
\ee
The cardinalities of the subsubsets are $a_n=\#\blacb_n$, $b_n=\#\bmub_n=\#\bxi_n$, $n=2,4$.

The  next step is to   express the functions $r_k$ in \eqref{Resh-SP-22} through the Bethe
equations:
\be{ru2}
 r_1(\blab_2)=\frac{f(\blab_2,\blab_1)f(\blab_2,\blab_3)f(\blab_2,\blab_4)}
 {f(\blab_1,\blab_2)f(\blab_3,\blab_2)f(\blab_4,\blab_2)}\;f(\bmub,\blab_2),
 \ee
\be{rv2}
r_3(\bmub_2)=\frac{f(\bmub_1,\bmub_2)f(\bmub_3,\bmub_2)f(\bmub_4,\bmub_2)}
{f(\bmub_2,\bmub_1)f(\bmub_2,\bmub_3)f(\bmub_2,\bmub_4)}f(\bmub_2,\blab),
\ee
\be{ru4}
 r_1(\blac_4)=\frac{f(\blac_4,\blac_1)f(\blac_4,\blac_2)f(\blac_4,\blac_3)}
 {f(\blac_1,\blac_4)f(\blac_2,\blac_4)f(\blac_3,\blac_4)}
 f(\bxi,\blac_4),
\ee
\be{rv4}
r_3(\bxi_4)=\frac{f(\bxi_1,\bxi_4)f(\bxi_2,\bxi_4)f(\bxi_3,\bxi_4)}
{f(\bxi_4,\bxi_1)f(\bxi_4,\bxi_2)f(\bxi_4,\bxi_3)}f(\bxi_4,\blac).
\ee

{\sl Remark 1.} We can explain now why we kept the notation $r_1(u)$ in spite of $r_1(u)=1$ in the TCBG model. This function
shows explicitly, to which subsubsets we should apply the Bethe equations in \eqref{Resh-SP-22}. In our case these are the
subsubsets $\blab_2$ and $\blac_4$.  If we had set $r_1(u)=1$ in \eqref{Resh-SP-22}, then we would have much more freedom
and we could use the Bethe equations for other subsubsets, which is inappropriate.

{\sl Remark 2.}
Note that we have used the Bethe equations for the product $r_3(\bxi_4)$ in spite of the set $\bxi$ contains the parameter $w$.
In fact, $w\in\bxi_{\so}$, hence,
$w\notin\bxi_{4}$, therefore we can use the Bethe equations for the product $r_3(\bxi_4)$. In the r.h.s. of \eqref{rv4} we have $w$ as the argument
of the $f$ functions, but this function goes to $1$ as $w\to+i\infty$.

Equations \eqref{Resh-SP-11}, \eqref{Resh-SP-22}, and \eqref{ru2}--\eqref{rv4} should be substituted into \eqref{Mat-el-12}. It leads
us to the following representation:
\begin{equation}\label{Mat-el2}
{\sf M}_{a,b}^{(3,2)}=\lim_{w\to+i\infty} e^{iwx} \frac{w}{c}\;\sum_{w\in\{\bxi_1,\bxi_3\}}
\frac{\ell_1(\blac_{2})\ell_1(\blac_{3})\ell_3(\bmub_{1})\ell_3(\bmub_{4})}{\ell_1(\blab_{2})\ell_1(\blab_{3})\ell_3(\bxi_{1})\ell_3(\bxi_{4})}\;
 F^C_{uu}\;F^C_{vv}\;F^C_{vu}
\;F^B_{uu}\;F^B_{vv}\;F^B_{vu}\;  \mathcal{Z}.
\end{equation}
Here the sum is taken over partitions of every set of Bethe parameters into four subsets
\be{subsubsets}
\begin{aligned}
\blacb&\Rightarrow\{\blacb_1,\blacb_2,\blacb_3,\blacb_4\},\\
\bxi&\Rightarrow\{\bxi_1,\bxi_2,\bxi_3,\bxi_4\},\\
\bmub&\Rightarrow\{\bmub_1,\bmub_2,\bmub_3,\bmub_4\},
\end{aligned}
\ee
with the restriction $w\in\{\bxi_1,\bxi_3\}$ that is shown explicitly by the subscript of the sum.
The factor $\mathcal{Z}$ in \eqref{Mat-el2} is the product of four highest coefficients
\be{Z}
\mathcal{Z}=Z_{a_3,b_1}(\blac_3;\blab_3|\bxi_1;\bmub_1) Z_{a_1,b_3}(\blab_1;\blac_1|\bmub_3;\bxi_3)\num
Z_{a_4,b_2}(\blac_4;\blab_4|\bxi_2;\bmub_2) Z_{a_2,b_4}(\blab_2;\blac_2|\bmub_4;\bxi_4).
\ee
Other factors in \eqref{Mat-el2} denoted by $F$ with different subscripts and superscripts are
products of $f$ functions:
\be{FCuu}
F^C_{uu}=f(\blac_4,\blac_1)
f(\blac_3,\blac_2)
f(\blac_4,\blac_2)
f(\blac_4,\blac_3)
f(\blac_1,\blac_2)
f(\blac_1,\blac_3),
\ee
\be{FBuu}
F^B_{uu}=f(\blab_1,\blab_4)
f(\blab_2,\blab_3)
f(\blab_2,\blab_1)
f(\blab_2,\blab_1)
f(\blab_3,\blab_4)
f(\blab_3,\blab_4),
\ee
\be{FCvv}
F^C_{vv}=f(\bxi_1,\bxi_4)
f(\bxi_2,\bxi_3)
f(\bxi_2,\bxi_1)
f(\bxi_2,\bxi_4)
f(\bxi_3,\bxi_1)
f(\bxi_3,\bxi_4),
\ee
\be{FBvv}
F^B_{vv}=f(\bmub_4,\bmub_1)
f(\bmub_3,\bmub_2)
f(\bmub_1,\bmub_3)
f(\bmub_4,\bmub_3)
f(\bmub_1,\bmub_2)
f(\bmub_4,\bmub_2),
\ee
\be{FCvu}
F^C_{vu}=f(\bxi_1,\blac_4)f(\bxi_4,\blac_4)f(\bxi_1,\blac_1)f(\bxi_4,\blac_1)
f(\bxi_3,\blac_4)f(\bxi_4,\blac_3),
\ee
\be{FBvu}
F^B_{vu}=f(\bmub_3,\blab_3)f(\bmub_2,\blab_2)f(\bmub_3,\blab_2)f(\bmub_2,\blab_3)
f(\bmub_1,\blab_2)f(\bmub_2,\blab_1).
\ee

It remains to combine the subsubsets into new groups:
\be{comb}
\begin{aligned}
&\{\blacb_1,\blacb_4\}=\blacb_{\rm i},\\
&\{\bxi_1,\bxi_4\}=\bxi_{\rm i},\\
&\{\bmub_1,\bmub_4\}=\bmub_{\rm i},
\end{aligned}
\qquad
\begin{aligned}
&\{\blacb_2,\blacb_3\}=\blacb_{\rm ii},\\
&\{\bxi_2,\bxi_3\}=\bxi_{\rm ii},\\
&\{\bmub_2,\bmub_3\}=\bmub_{\rm ii}.
\end{aligned}
\ee
Then we obtain
\begin{multline}\label{Mat-el4-12}
{\sf M}_{a,b}^{(3,2)}=\lim_{w\to+i\infty} e^{iwx} \frac{w}{c}\;\sum
\frac{\ell_1(\blac_{\rm ii})\ell_3(\bmub_{\rm i})}{\ell_1(\blab_{\rm ii})\ell_3(\bxi_{\rm i})}\;
f(\blac_{\rm i},\blac_{\rm ii})f(\blab_{\rm ii},\blab_{\rm i})f(\bxi_{\rm ii},\bxi_{\rm i})f(\bmub_{\rm i},\bmub_{\rm ii})
\\
\times f(\bxi_{\rm i},\blac_{\rm i})f(\bmub_{\rm ii},\blab_{\rm ii})\;G_1(\blac_{\rm i},\blab_{\rm i};\bxi_{\rm ii},\bmub_{\rm ii})\,
 G_2(\blac_{\rm ii},\blab_{\rm ii};\bxi_{\rm i},\bmub_{\rm i}),
\end{multline}
where the sum is taken over partitions $\blacb\Rightarrow\{\blacb_{\rm i},\blacb_{\rm ii}\}$, $\bxi\Rightarrow\{\bxi_{\rm i},\bxi_{\rm ii}\}$,
and $\bmub\Rightarrow\{\bmub_{\rm i},\bmub_{\rm ii}\}$.
The functions $G_1$ and $G_2$ in their turn are given as the sums over partitions of the subsets above into subsubsets:
 \begin{equation}\label{G1-12}
 G_1(\blac_{\rm i},\blab_{\rm i};\bxi_{\rm ii},\bmub_{\rm ii})
 =\sum_{w\notin\bxi_{2}}
   W\begin{pmatrix}
  \blac_{1},\blab_{1},&\bxi_{2},\bmub_{2}\\
    \blac_{4},\blab_{4},&\bxi_{3},\bmub_{3}
  \end{pmatrix},
 \end{equation}
 and
 \begin{equation}\label{G2-12}
 G_2(\blac_{\rm ii},\blab_{\rm ii};\bxi_{\rm i},\bmub_{\rm i})=\sum_{w\notin\bxi_{4}}
  W\begin{pmatrix}
  \blac_{2},\blab_{2},&\bxi_{1},\bmub_{1}\\
    \blac_{3},\blab_{3},&\bxi_{4},\bmub_{4}
  \end{pmatrix},
 \end{equation}
where $W$ is defined by \eqref{Resh-SP-eig2}.

In \eqref{G1-12} and \eqref{G2-12} we have the sums over partitions with the restrictions indicated explicitly by the subscripts of the sums.
These restrictions appear due to the original condition $w\in\bxi_{\so}$, that implies $w\in\{\bxi_1,\bxi_3\}$.  It is easy to see, however,
that actually one of these sums has no  restriction.

Indeed, suppose that $w\in\bxi_1$. Then the set $\bxi_{\rm ii}=\{\bxi_2,\bxi_3\}$ does not contain the parameter $w$. Hence, no restriction is imposed on the
sum \eqref{G1-12}. Then due to \eqref{Wpart} we conclude that $G_1=0$, unless $\blacb_{\rm i}=\bmub_{\rm ii}=\bxi_{\rm ii}=\emptyset$.
Then $\blacb_{\rm ii}=\blacb$, $\bmub_{\rm i}=\bmub$, and $\bxi_{\rm i}=\bxi$.

Similarly, if $w\in\bxi_3$, then the set $\bxi_{\rm i}=\{\bxi_1,\bxi_4\}$ does not contain the parameter $w$, and therefore we have no restrictions
in the sum \eqref{G2-12}. Hence,
$G_2=0$,  unless $\blacb_{\rm ii}=\bmub_{\rm i} =\bxi_{\rm i}=\emptyset$. Then
$\blacb_{\rm i}=\blacb$, $\bmub_{\rm ii}=\bmub$, and $\bxi_{\rm ii}=\bxi$. We arrive at the following representation
\be{M-LL}
{\sf M}_{a,b}^{(3,2)}=\lim_{w\to+i\infty} e^{iwx}\frac{w}{c}\;\left(\Omega_1\frac{\ell_1(\blac)\ell_3(\bmub)}{\ell_1(\blab)\ell_3(\bxi)}+\Omega_2\right),
\ee
where
 \begin{equation}\label{Om0}
 \Omega_1
 =\sum_{w\in\bxi_{1}}
   W\begin{pmatrix}
  \blac_{2},\blab_{2},&\bxi_{1},\bmub_{1}\\
    \blac_{3},\blab_{3},&\bxi_{4},\bmub_{4}
  \end{pmatrix},\qquad
 \Omega_2=\sum_{w\in\bxi_{3}}
  W\begin{pmatrix}
  \blac_{1},\blab_{1},&\bxi_{2},\bmub_{2}\\
    \blac_{4},\blab_{4},&\bxi_{3},\bmub_{3}
  \end{pmatrix}.
 \end{equation}
 Relabeling the subsets we obtain
 \begin{equation}\label{Om1}
 \Omega_1
 =\sum_{w\in\bxi_{\so}}
   W\begin{pmatrix}
  \blac_{\so},\blab_{\so},&\bxi_{\so},\bmub_{\so}\\
    \blac_{\st},\blab_{\st},&\bxi_{\st},\bmub_{\st}
  \end{pmatrix},\qquad
 \Omega_2=\sum_{w\in\bxi_{\st}}
  W\begin{pmatrix}
  \blac_{\so},\blab_{\so},&\bxi_{\so},\bmub_{\so}\\
    \blac_{\st},\blab_{\st},&\bxi_{\st},\bmub_{\st}
  \end{pmatrix}.
 \end{equation}
It is clear that in the limit $w\to+i\infty$ the coefficients $\Omega_{1}$ and $\Omega_{2}$ respectively coincide with
\eqref{W-T32} and \eqref{W-T32-d}. Indeed, these coefficients are rational functions of their arguments, therefore it is not
important how the parameter $w$ approaches infinity. Thus, we obtain
\be{M-OmOm}
{\sf M}_{a,b}^{(3,2)}=\lim_{w\to+i\infty} e^{iwx}\left(\frac{\ell_1(\blac)\ell_3(\bmub)}{\ell_1(\blab)\ell_3(\bxi)}-1\right)
\lim_{w\to+i\infty} \frac{w}{c}\;\Omega_1=
\frac{\ell_3(\bmub)}{\ell_3(\bmuc)}\;\mathfrak{F}_{a,b}^{(3,2)}(\blac,\bmuc;\blab,\bmub),
\ee
where we have used $\ell_1(u)=1$. Thus, we have calculated the form factor of the right partial zero mode $\ZMRp{32}$ and
using now \eqref{0Ti3-xpsi} we arrive at \eqref{FF-psik-res} for $k=2$.

\section*{Conclusion}

In this paper we considered form factors of local operators in the TCBG model. We have shown that they can be reduced to the
universal form factors of the monodromy matrix entries. The latter were calculated in
\cite{BelPRS12b,BelPRS13a,PakRS14b,PakRS14c,PakRS15a}, where determinant representations were found. Determinant formulas
for form factors allow one to study the problem of correlation functions. It was done already for the model of the one-component
Bose gas \cite{KitKMST12,CauCS07,PanC14}. We hope that the formulas obtained in the present paper will play the same role in studying the TCBG model.
Indeed, knowing form factors one can attack the problem of local operators correlation functions. It gives a possibility to compare theoretical predictions with the experimental results obtained for strongly correlated quantum systems (see e.g. \cite{MorSKE03,LabHHPRP03,ParatALL04,KinWW04,PolRD04}).

We also have seen that the zero modes method used in \cite{PakRS15c} for the evaluation of the local operators form factors
can be adapted to the case of TCBG model. In spite of these modifications the final results are very similar
to the results of \cite{PakRS15c}. At least the most essential parts of both results are given by the universal form factors
of the monodromy matrix entries. It may happen that such  universal coefficients also take place for other integrable models, in
particular, for the models described by the $q$-deformed (trigonometric) $R$-matrix. The study of this question is now in
progress.

\section*{Acknowledgements}
It is a great pleasure for us to thank J.-S.~Caux who attracted our attention to the problem considered in the present paper.
The work of S.P. was supported in part by RFBR-14-01-00474-a.
N.A.S. was  supported by the Program of RAS  ``Nonlinear Dynamics in Mathematics and Physics'',
RFBR-14-01-00860-a, RFBR-13-01-12405-ofi-m2.


\appendix

\section{Proof of lemma~\ref{Lem-FF32}\label{A-PL}}

The strategy for the proof of  lemma~\ref{Lem-FF32} is as follows.  We consider  an auxiliary model solvable by the
algebraic Bethe ansatz and possessing the $R$-matrix \eqref{R-mat}. We present the monodromy matrix of this auxiliary model
in the form \eqref{T-TT} and assume that the total and partial monodromy matrices have the following asymptotic expansions:
\be{Azero-modes}
\begin{aligned}
&T(u)=\mathbf{1}+ \sum_{n=0}^\infty T[n]\,\left(\frac cu\right)^{n+1},\\
&T^{(l)}(u)=\mathbf{1}+ \sum_{n=0}^\infty T^{(l)}[n]\,\left(\frac cu\right)^{n+1},\qquad l=1,2.
\end{aligned}
\ee
Respectively the total and partial zero modes are defined as
\be{A0-testij}
\begin{aligned}
&T_{ij}[0]=\lim_{|u|\to\infty}\frac uc(T_{ij}(u)-\delta_{ij}),\\
&T^{(l)}_{ij}[0]=\lim_{|u|\to\infty}\frac uc(T^{(l)}_{ij}(u)-\delta_{ij}), \qquad l=1,2,
\end{aligned}
\qquad i,j=1,2,3.
\ee
Such a type of models was considered in \cite{PakRS15c}, where the form factors of all partial zero modes were
computed. In particular, using the notation of section \ref{S-MR}
\be{Univ-23}
{\sf \widetilde M}_{a,b}^{(3,2)}(\blac,\bmuc;\blab,\bmub)=\left(\frac{\ell_1(\blac)\ell_3(\bmub)}{\ell_1(\blab)\ell_3(\bmuc)}-1\right)
\mathfrak{F}_{a,b}^{(3,2)}(\blac,\bmuc;\blab,\bmub),
\ee
where
\be{denote-32}
{\sf \widetilde M}_{a,b}^{(3,2)}\equiv {\sf \widetilde M}_{a,b}^{(3,2)}(\blac,\bmuc;\blab,\bmub)=
\mathbb{C}_{a,b-1}(\blac;\bmuc) T^{(1)}_{32}[0] \mathbb{B}_{a,b}(\blab;\bmub).
\ee
Recall also that $\mathfrak{F}_{a,b}^{(3,2)}(\blac,\bmuc;\blab,\bmub)$ in \eqref{Univ-23} is the universal form factor of the operator
$T_{32}(z)$. The main property of the universal form factor is that it  does not depend on the model under consideration. Thus, it
is the same for all the models possessing the $R$-matrix \eqref{R-mat}.

Now we should reproduce this result by the straightforward method used in section~\ref{SS-PSI}. Namely, using the action formula
\be{actdBVa}
\mathbb{C}^{(1)}_{a,b}(\bla;\bmu)T^{(1)}_{32}[0]=\lim_{|w|\to\infty} \frac wc\;\mathbb{C}^{(1)}_{a,b+1}(\bla;\{w,\bmu\}),
\ee
and  the scalar product \eqref{Resh-SP},
we will see that  form factor \eqref{denote-32} reduces  to the sum
\eqref{W-T32}. Comparing the results we obtain the statement of lemma~\ref{Lem-FF32}.

Due to \eqref{CV-CV-eig} and \eqref{actdBVa} the action of $T^{(1)}_{32}[0]$ on a total dual on-shell
Bethe vector  is
\begin{multline}\label{act-BV-BV12}
\mathbb{C}_{a,b-1}(\blac;\bmuc) T^{(1)}_{32}[0]=\lim_{|w|\to\infty} \frac{w}{c}\;\sum
\frac{\ell_{3}(\bmuc_{\st})}{\ell_{1}(\blac_{\so})} f(\blac_{\so},\blac_{\st})f(\bmuc_{\st},\bmuc_{\so})
f(\bmuc_{\so},\blac_{\so})\\
\times\mathbb{C}_{a_{\so},b_{\so}}^{(1)}(\blac_{\so};\{w,\bmuc_{\so}\}) \mathbb{C}_{a_{\st},b_{\st}}^{(2)}(\blac_{\st};\bmuc_{\st}).
\end{multline}
Let $\{w,\bmuc\}=\bxi$. Then
\begin{multline}\label{Mat-el-12}
{\sf \widetilde M}_{a,b}^{(3,2)}=\lim_{|w|\to\infty} \frac{w}{c}\;\sum_{w\in\bxi_{\so}}
\frac{\ell_1(\blac_{\st})\ell_3(\bmub_{\st})}{\ell_1(\blab_{\so})\ell_3(\bxi_{\so})}\;
 f(\blac_{\so},\blac_{\st})f(\blab_{\so},\blab_{\st})f(\bxi_{\st},\bxi_{\so})f(\bmub_{\st},\bmub_{\so})
\\
\times f(\bmub_{\so},\blab_{\so})f(\bxi_{\so},\blac_{\so})\;
\mathbb{C}_{a_{\so},b_{\so}+1}^{(1)}(\blac_{\so};\bxi_{\so})\mathbb{B}_{a_{\so},b_{\so}+1}^{(1)}(\blab_{\so};\bmub_{\so}) \cdot \mathbb{C}_{a_{\st},b_{\st}}^{(2)}(\blac_{\st};\bxi_{\st})\mathbb{B}_{a_{\st},b_{\st}}^{(2)}(\blab_{\st};\bmub_{\st}).
\end{multline}
The sum in \eqref{Mat-el-12} is taken over the partitions
$\blacb\Rightarrow\{\blacb_{\so},\blacb_{\st}\}$, $\bxi\Rightarrow\{\bxi_{\so},\bxi_{\st}\}$,
and $\bxi\Rightarrow\{\bxi_{\so},\bxi_{\st}\}$ with a restriction
$w\in\bxi_{\so}$.  We also have set  $\bxi_{\so}=\{w,\bmuc_{\so}\}$ and  $\bxi_{\st}=\bmuc_{\st}$.

Comparing \eqref{Mat-el-12} with \eqref{act-BV-BV12} we see that we replaced the product
$\ell_3(\bmuc_{\so})$ by $\ell_3(\bxi_{\so})$. We can do it, because  $\ell_3(w)\to 1$ as $|w|\to\infty$ due to the
asymptotic expansion \eqref{Azero-modes}. Similarly to \eqref{Mat-el-32}
we also replaced the products $f(\bmuc_{\st},\bmuc_{\so})$ and $f(\bmuc_{\so},\blac_{\so})$ by
$f(\bxi_{\st},\bxi_{\so})$ and $f(\bxi_{\so},\blac_{\so})$, because the $f$ function   goes to $1$ if one of its
arguments goes to infinity.

It remains now to repeat the derivation of section~\ref{SS-PSI}. All transforms of \eqref{Mat-el-12} are {\it exactly the same}
as we did for equation \eqref{Mat-el-32}. Therefore we  give  here only the final result, which is an analog of \eqref{M-OmOm}
\be{AM-OmOm}
{\sf\widetilde M}_{a,b}^{(3,2)}=\lim_{|w|\to\infty} \left(\frac{\ell_1(\blac)\ell_3(\bmub)}{\ell_1(\blab)\ell_3(\bxi)}-1\right)
\frac{w}{c}\;\Omega_1,
\ee
where $\Omega_1$ is given by the sum \eqref{Om1}. In distinction to \eqref{M-OmOm} now we have $\ell_3(w)\to 1$ as $|w|\to\infty$,
and hence,
\be{AM-OmOm-1}
{\sf\widetilde M}_{a,b}^{(3,2)}= \left(\frac{\ell_1(\blac)\ell_3(\bmub)}{\ell_1(\blab)\ell_3(\bmuc)}-1\right)
\lim_{|w|\to\infty}\frac{w}{c}\;\Omega_1.
\ee
Comparing \eqref{AM-OmOm-1} and \eqref{Univ-23} we see that
\be{Om-FF}
\lim_{|w|\to\infty}\frac{w}{c}\;\Omega_1=
\mathfrak{F}_{a,b}^{(3,2)}(\blac,\bmuc;\blab,\bmub),
\ee
and this is exactly the statement of lemma~\ref{Lem-FF32}.

\end{document}